\documentclass[a4paper,11pt,full]{article}
\usepackage{nicefrac}
\usepackage{pos}

\title{Search for continuous phase transitions in 5D pure SU(2) lattice gauge
theory}
\author*[a]{José Matos}
\affiliation[a]{Centro de Física das Universidades do Porto, University of Porto,\\ 4169-007, Porto, Portugal}
\emailAdd{up201506054@edu.fc.up.pt}
\abstract{
The Renormalization Group (RG) is one of the central and modern techniques
in quantum field theory. Indeed, quantum field theories can be understood
as flows between fixed points of the RG flow, which represent Conformal Field Theories
(CFT's). Hence, the search and classification of yet unknown
non-trivial CFT's is a legitimate endeavor. Analytical considerations
point to the existence of such a fixed point in pure SU(2) Yang-Mills
fields in 5D. This issue has already been addressed, although inconclusively.
We search for this putative fixed point using lattice Monte Carlo methods. An extended discussion can be found in \cite{Florio:2021uoz}. }

\FullConference{%
 The 38th International Symposium on Lattice Field Theory, LATTICE2021
  26th-30th July, 2021
  Zoom/Gather@Massachusetts Institute of Technology
}

\begin{document}

\maketitle

\section{Introduction}
A widely used framework to classify the set of consistent field theories
is the Renormalization Group (RG), see \cite{Wilson:1973jj} for
a review. In this framework, there is a special set of theories signaled by the fix points of the the RG flow. We investigate the existence of such a fixed point in the 5-dimensional
pure SU(2) Yang-Mills theory. The existence of a non-trivial fixed
point in this model was first proposed in\cite{Peskin:1980ay}, in the context
of the $\varepsilon$-expansion in $4+\varepsilon$ and subsequently
studied numerically in \cite{Kawai:1992um}. If this putative fixed point exists,
it would be one of the few examples of non-trivial conformal field
theory (CFT's) in $D>4$. 

This model is particularly interesting due to the non-renormalizability
of the theory in $5$ dimensions, which excludes it from being studied perturbatively, similarly to the non-linear
sigma model which is non-renormalizable in $3$ dimensions. The latter, however, has
a non-trivial fixed point \cite{Wilson:1973jj,Polyakov:1975rr,Bardeen:1976zh,Arefeva:1978fj}.
The existence of non-trivial fixed points in non-renormalizable theories
is also the main idea behind the "asymptotic safety" program,
see \cite{Eichhorn:2018yfc} for a review. 

The study of this non-perturbative effects requires the use of non-perturbative methods. We elected to use Monte-Carlo methods to study a lattice version of the theory, in which fixed points of the RG flow are identified with continuous phase transitions.
Based on the results from the $\varepsilon$-expansion \cite{Peskin:1980ay},
we expect the critical surface to have co-dimension one, which means
that we should have to tune a single parameter (or in the unlikely
situation that one of the parameters flows parallel to the critical
surface, two) to hit the critical surface. Previous results \cite{Kawai:1992um,Creutz:1979dw}
rule out this possibility, as they only identified first order phase transitions.

We will follow a similar procedure to that presented on \cite{Kawai:1992um,Creutz:1979dw},
but we will take advantage of the freedom to chose different
lattice actions which the same naive continuum limit but with different projections
into the continuum operators. In particular, we will focus
on lattice actions with two free parameters in order to increase
our chances of hitting the critical surface. This work is presented in detail in \cite{Florio:2021uoz}.

\section{Lattice Actions and Monte Carlo}

We used two simple extensions of the usual Wilson action, introduced
in \cite{Wilson:1973jj}, since it preserves local gauge invariance. This is achieved at the expense of introducing
link variables $U_{\mu}^{\Re},$ belonging to any representation of
$SU\left(2\right)$. In the pure version (i.e. without fermions) the
only gauge invariant objects are traces of ordered products of the link variables around
closed loops. These loops are usually referred
to as Wilson loops. The simplest Wilson loop is a rectangular loop
contained in a plane defined by two directions $\mu$, $\nu$ and
with dimension $I$ and $J$, respectively. We will denote these
loops as $\square_{\mu\nu,I\times J}^{R}$. In this work we will only
consider isotropic systems such that the directional indexes $\mu$
and $\nu$ can be dropped. With these conventions, the standard Wilson
action takes the form
\[
S=\sum_{\square}\frac{\beta}{2}\operatorname{Tr}\left(1-\square_{1\times1}^{f}\right),
\]
 where the sum is performed over all distinct $1\times1$ Wilson loops
and the link variables are in the fundamental representation. 

We will study two different variations of this action. The first,
already studied in \cite{Kawai:1992um}, considers an extra term
where the trace of the link variables is taken in the adjoint representation\footnote{The numeric factors are added to normalize the traces as since the
trace of the identity in the fundamental representation is 2 and in
the adjoint representation is 3. } 
\begin{equation}
S_{1\times1}^{f,a}=\sum_{\square_{1\times1}}\frac{\beta_{f}}{2}\operatorname{Tr}\left(1-\square_{1\times1}^{f}\right)+\frac{\beta_{a}}{3}\operatorname{Tr}\left(1-\square_{1\times1}^{a}\right).\label{eq:fundamental + adjoint action}
\end{equation}
We selected his extension as a starting point, since in \cite{Kawai:1992um}
the authors suggest that a continuous phase transition may exist in this model for negative
enough values of $\beta_{a}$.
The second extension uses loops of
different sizes in the fundamental representation 
\begin{equation}
S_{1\times1,2\times2}^{f}=\sum_{\square_{1\times1}}\frac{\beta_{1}}{2}\operatorname{Tr}\left(1-\square_{1\times1}^{f}\right)+\sum_{\square_{2\times2}}\frac{\beta_{2}}{2}\operatorname{Tr}\left(1-\square_{2\times2}^{f}\right),\label{eq:2 loops action}
\end{equation}
and it was chosen to explore a different region of the coupling space. 

It is useful to define the lattice average of the trace of the Wilson
loops
\[
W_{I\times J}^{\Re}=1-\frac{\sum_{\square\times J}\operatorname{Tr}\left(\square_{I\times J}^{\mathfrak{R}}\right)}{N_{\text{loops }}d(\mathfrak{R})},
\]
where $N_{\text{loops }}$ is the number of loops of the given type
and $d(\mathfrak{R})$ is the dimension of the representation. In this notation, the extended actions take a particularly simple form
\[
\begin{aligned}S_{1\times1}^{f,a} & =N_{\text{loops }}\left(\beta_{f}W_{1\times1}^{f}+\beta_{a}W_{1\times1}^{a}\right),\\
S_{1\times1,2\times2}^{f} & =N_{\text{loops }}\left(\beta_{1}W_{1\times1}^{f}+\beta_{2}W_{2\times2}^{f}\right).
\end{aligned}
\]

The standard Monte Carlo approach is based on generating large numbers of configurations to perform measurements of  some average in the canonical ensemble. In this context, we identify a phase
transition as a qualitative change in the dependence of this average
in the coupling being changed (e.g. the order parameter of the transition
should be zero for some values of $\beta$ and non-zero for others).
If we compute the average of the order parameter, this change is discontinuous for first order phase transition  and it is continuous for a continuous (second order) phase transition.

\begin{figure}[t]
\begin{centering}
\includegraphics[width=0.45\linewidth]{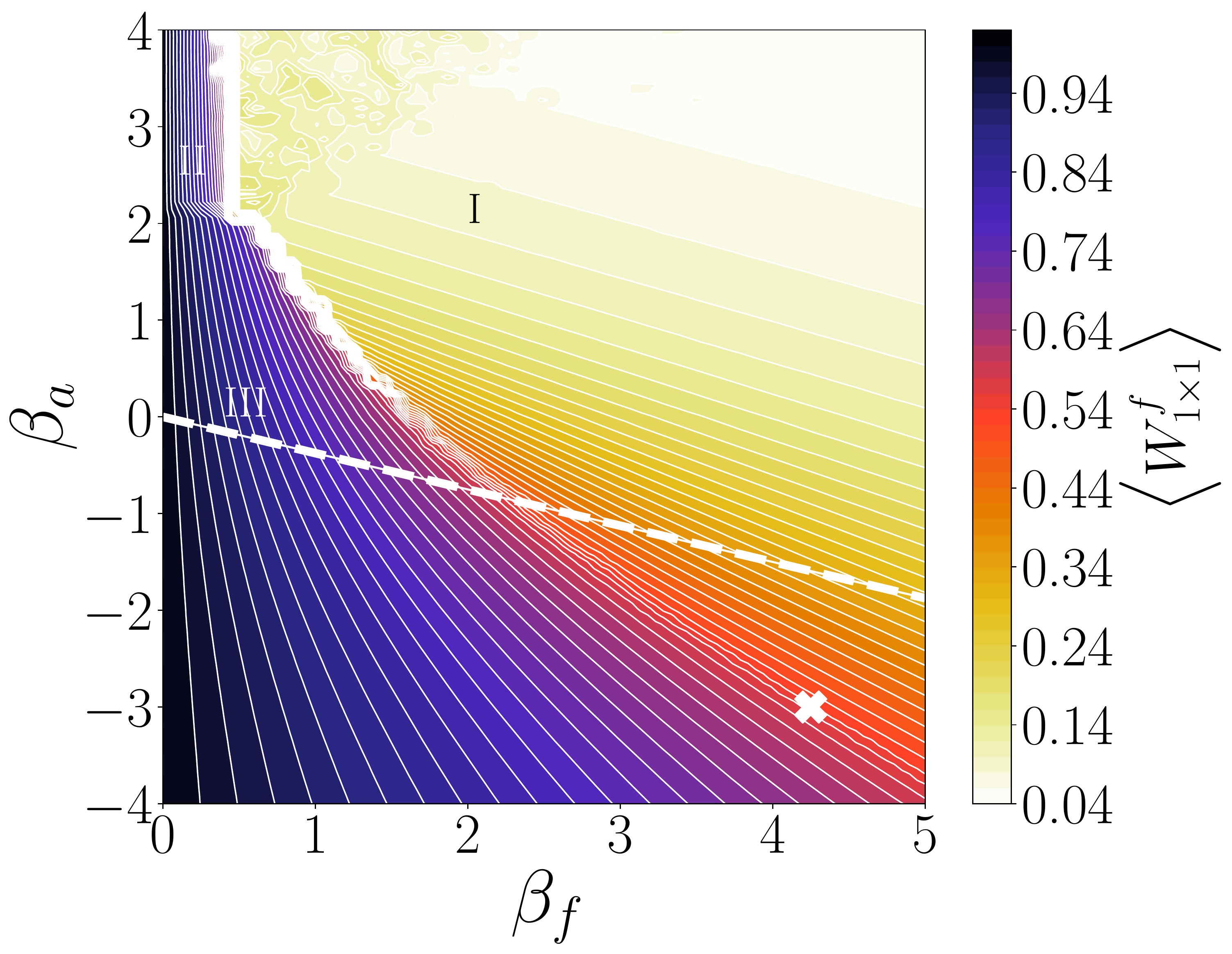}\includegraphics[width=0.45\linewidth]{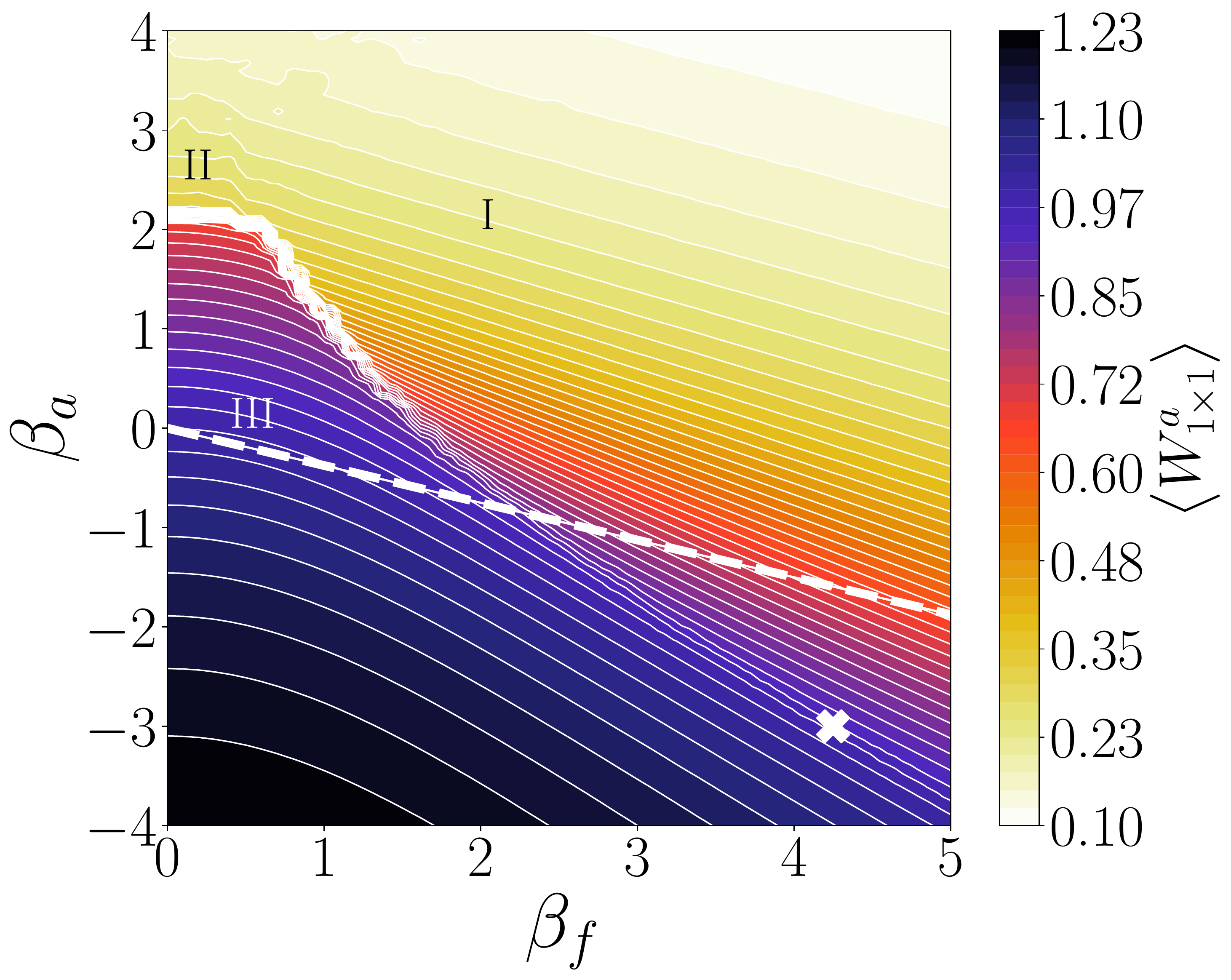}
\par\end{centering}
\caption{Survey of the parameter space, for $L=4$. We show $\left\langle W_{1\times1}^{f}\right\rangle _{\beta_{f},\beta_{a}}$
on the \textbf{left} and $\left\langle W_{1\times1}^{a}\right\rangle _{\beta_{f},\beta_{a}}$
on the \textbf{right} . The white cross signals some of the parameters analysed in
\cite{Kawai:1992um}. We identified three different \textquotedbl phases\textquotedbl ,
labeled by the roman-numerals. The thin white lines are level curves.
Discontinuities are identified when several white lines come together.
The dashed white line distinguishes regions with different signs of
the coupling in the naive continuum limit; below $\nicefrac{1}{g^{2}}<0$
and above $\nicefrac{1}{g^{2}}>0$. We used rectangular grid with
a spacing of $0.1$.\label{fig:PhaseSpaceScan}}
\end{figure}

For the models considered, this approach has two severe drawbacks. On
one hand, there is no simple order parameter whose change from
a non-zero value to zero can be used to identify a phase transition.
On the other hand, from the previous works in \cite{Kawai:1992um},
we need to distinguish very weak first order criticality
from true second order criticality, such that we are in the regime where the noise of the data may be bigger than the changes in the parameters.  We tackle these problems by introducing
a generalized ensemble that allows us to directly estimate the first
derivative of the entropy (i.e. the logarithm of the number
of configurations with a given value of $W_{1\times1}^{f}$) with respect to $W_{1\times1}^{f}$, which
enable us to identify the phase transition and its order. When the first derivative of the entropy is monotonic there are no phase transitions and when it is non-monotonic there is a first order phase transition. A second order phase transition occurs in the marginal case that separates the previous scenarios, when there is a point where the second derivative is zero. 

The idea behind our approach is replacing the canonical ensemble weight for a more adequate weight
\[
e^{-\beta WN}\to e^{-\left(\beta_{1}+\beta_{2}\frac{W}{N}\right)WN},
\]
where $N$ is the total number of loops and $W$ is the lattice average
of the trace of some Wilson loop. The probability density for a given
value of $W$ in this generalized ensemble is
\[
\rho\left(W\right)=e^{\mathbb{S}\left(W\right)-\left(\beta_{1}+\beta_{2}\frac{W}{N}\right)WN}\label{Probability distribution}.
\]
Notice that Monte Carlo simulations will generate an estimate for $\rho\left(W\right)$, such that we can always recover the entropy.

We can understand the advantage of this modification by recalling that Monte Carlo simulations usually sample the maximums of the probability distribution (Eq.\ref{Probability distribution}), and the number of local maximums of this distribution greatly impacts the
quality of the simulations. In particular, when there are two local
maximums, we usually observe hysteresis which can skew our measurements. It
can be shown that, by choosing a value of $\beta_{2}$ sufficiently
large, we can guarantee the existence of a single maximum, completely
avoiding the hysteresis problem and sampling even the thermodynamically
unstable region. With this method, we were able to obtain estimates
of the entropy, for all values of the action around the critical region.
This method is inspired by the usual multicanonical methods \cite{PhysRevE.74.046702,Berg:1992qua,PhysRevLett.71.211,PhysRevLett.63.1195},
in which the ensemble weight is chosen such that it cancels the entropy
and the resulting histogram, obtained in the Monte Carlo simulations,
is flat.

\section{Results }

In this section we will present the results obtained in this work.
We start by exploring the fundamental + adjoint action Eq.\ref{eq:fundamental + adjoint action},
with the aim of investigating the suggestion made in \cite{Kawai:1992um}
about the existence of second order criticality at $\beta_{a}\sim-6$.
We surveyed the parameter space for small lattice sizes
($L=4$) and identified the different phases of our model. Then,
we measure the first derivative of the entropy around the region
in which the first order phase transition becomes weaker, and perform
finite size scaling to check if our results hold in the infinite limit
size.

\begin{figure}[t]
\centering{}\includegraphics[width=0.50\columnwidth]{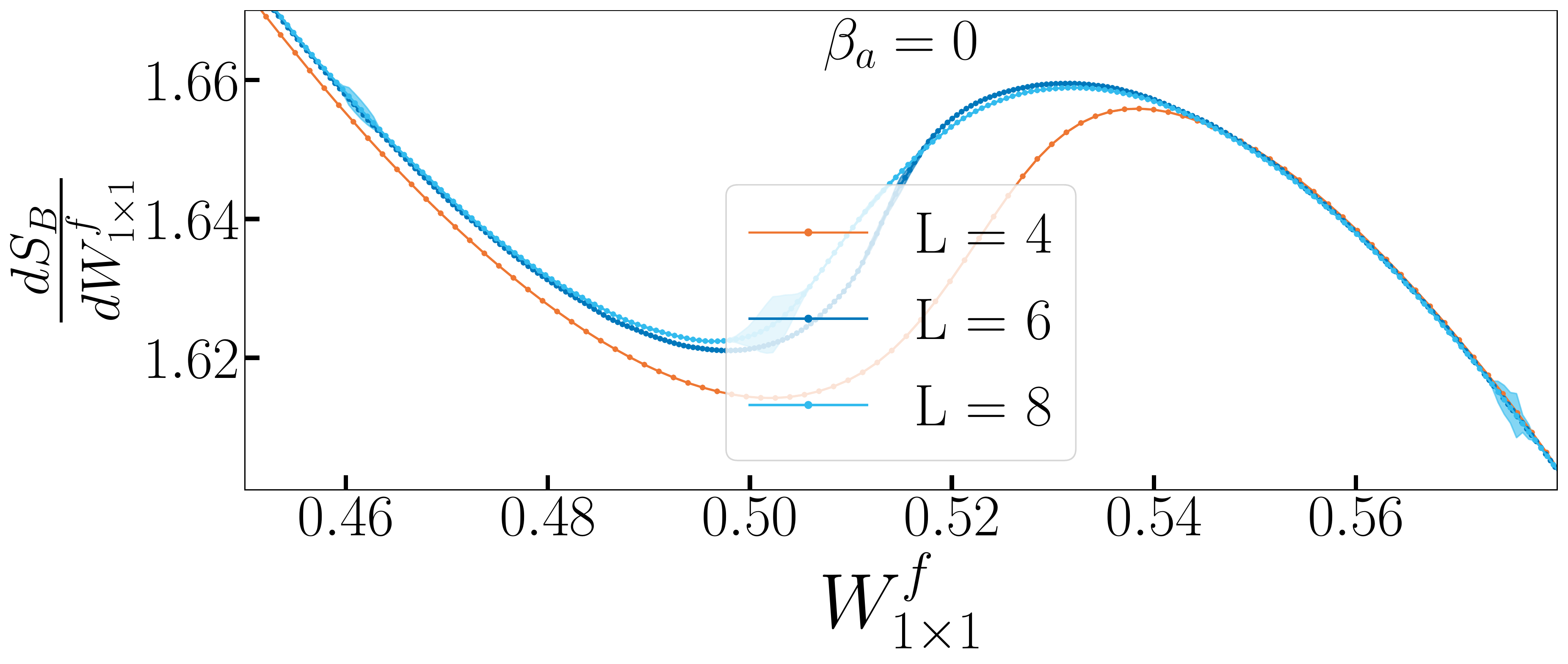}\includegraphics[width=0.5\columnwidth]{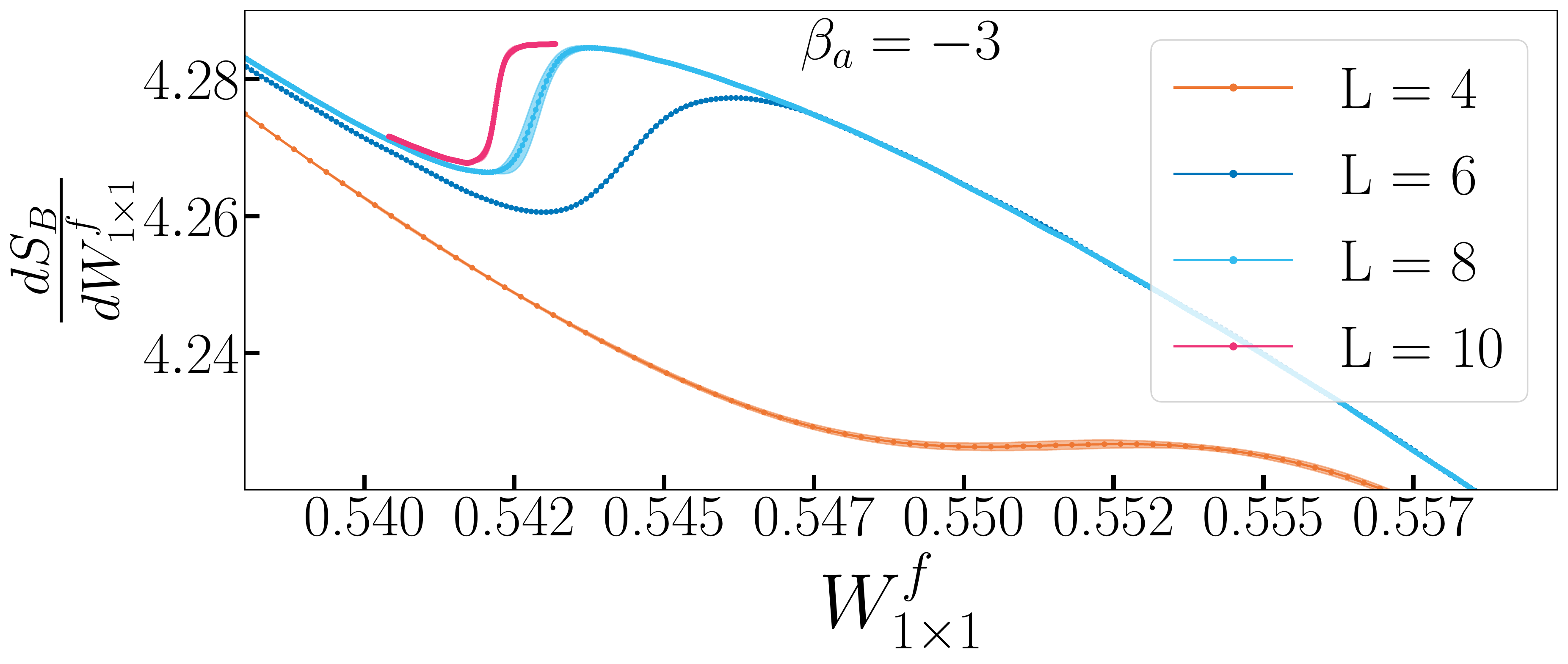}
\includegraphics[width=0.5\columnwidth]{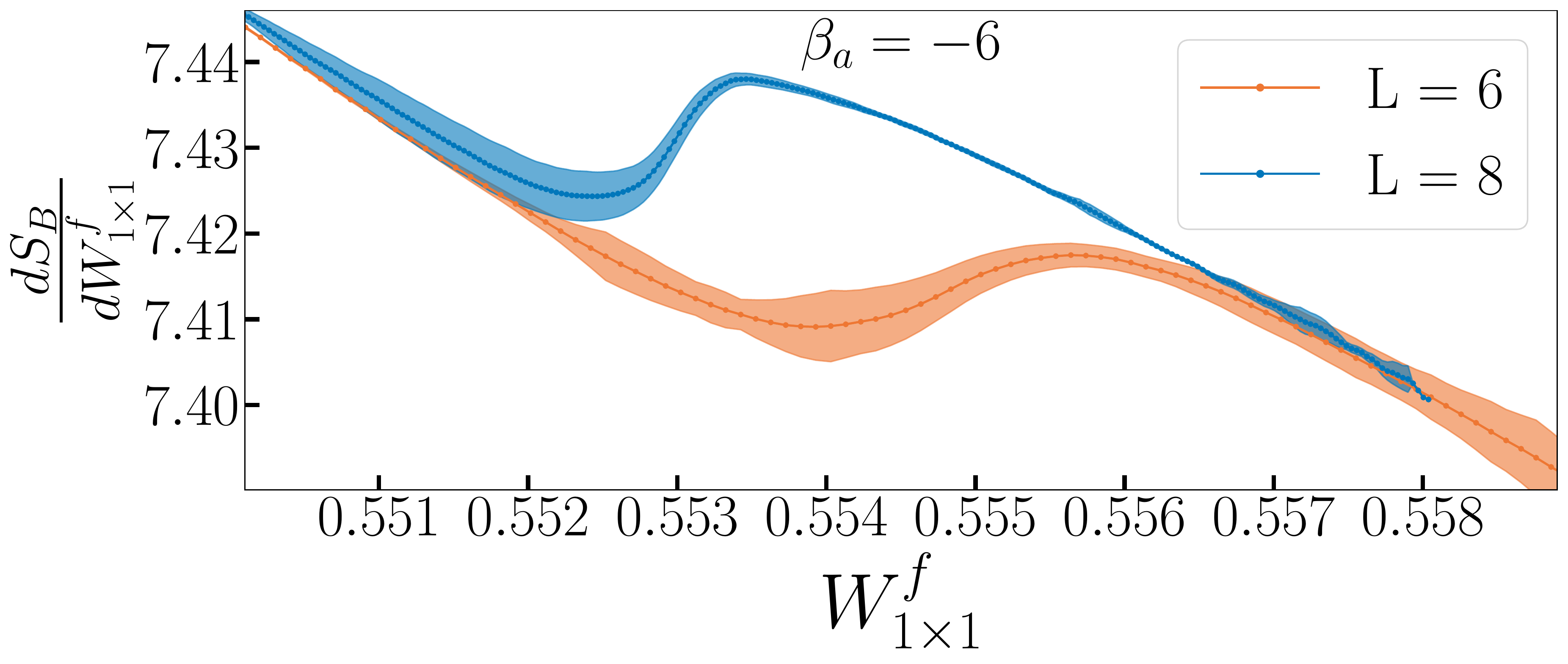}\caption{First derivative of the entropy with respect to $W_{1\times1}^{f}$,
as a function of $W_{1\times1}^{f}$, for different lattice sizes
and $\beta_{a}$. The colored region around the markers is an estimate
of the error based on independent measurements. \label{fig:First Derivative Adjoint}}
\end{figure}
\subsection{Fundamental + Adjoint Action}
In this extension, we identified several lines of first order criticality.
Although we could identify a curve over which first order criticality
becomes weaker, we show that the existence of second order criticality
is very unlikely. As mentioned, we started our search
by broadly exploring the phase diagram for small lattice sizes. The
results are shown in Fig.\ref{fig:PhaseSpaceScan} and suggest the
existence of 3 distinct phases with 3 curves of phase transitions separating
them. We studied each of these curves and found first order criticality
in all of them. However, in the boundary between phase I and phase
III, we observed a weakening of critical behavior as we decrease $\beta_{a}$.
This is the curve previously identified in \cite{Kawai:1992um}.
Besides the results presented in this report, we also performed measurements
of the "Creutz ratios" \cite{Creutz:1980wj} that suggest this line separates a confined phase from a deconfined phase. 

These results motivated a more complete study of this region. We studied in detail the phase transition over this line by measuring first derivative of the entropy in several points and studying how some of its characteristics change over this curve. The results for $\beta_{a}=0$, $\beta_{3}=-3$, and
$\beta_{a}=-6$ are presented in Fig.\ref{fig:First Derivative Adjoint}.
The two main features of this measurements are: 1) The width of the
thermodynamically unstable region decreases when $\beta_{a}$ decreases,
suggesting that for negative enough $\beta_{a}$, the first order phase
transition may vanish; 2) Increasing the system size not only does
not increase the width of the thermodynamically unstable region, but
it also makes it more unstable (the slope of the second derivative in the
thermodynamically unstable region increases). We verified this behavior
up to $\beta_{a}=-10$ and $L=8$. 

A second order phase transition requires that both the width of the thermodynamically unstable region 
and the maximum of the second derivative to go to zero.
Assuming the tendencies observed hold outside the range of system
sizes and couplings measured, we conclude that it is very unlikely that a second order phase transition
is presented in this extension.

\begin{figure}[t]
\begin{centering}
\includegraphics[width=0.45\linewidth]{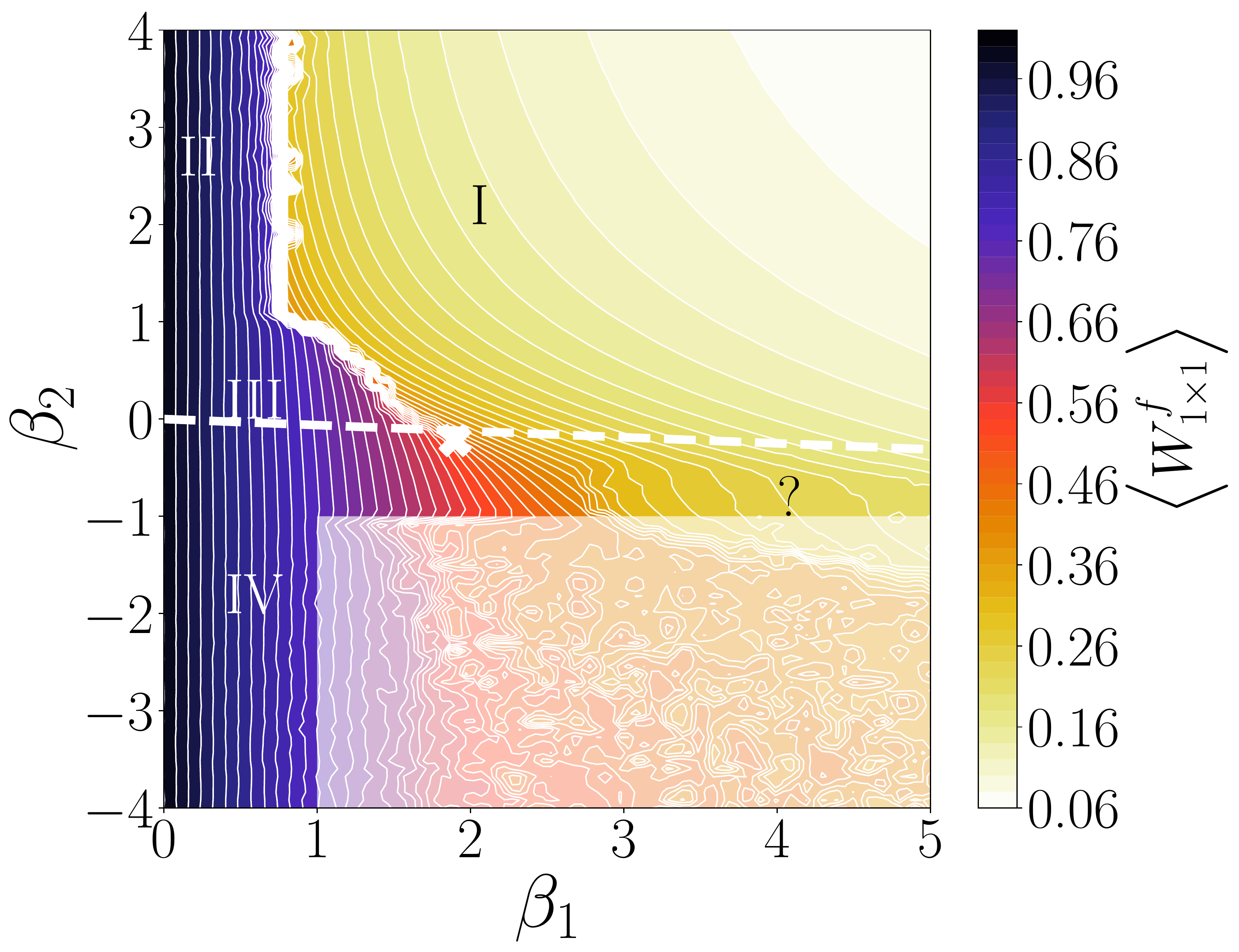}\includegraphics[width=0.45\linewidth]{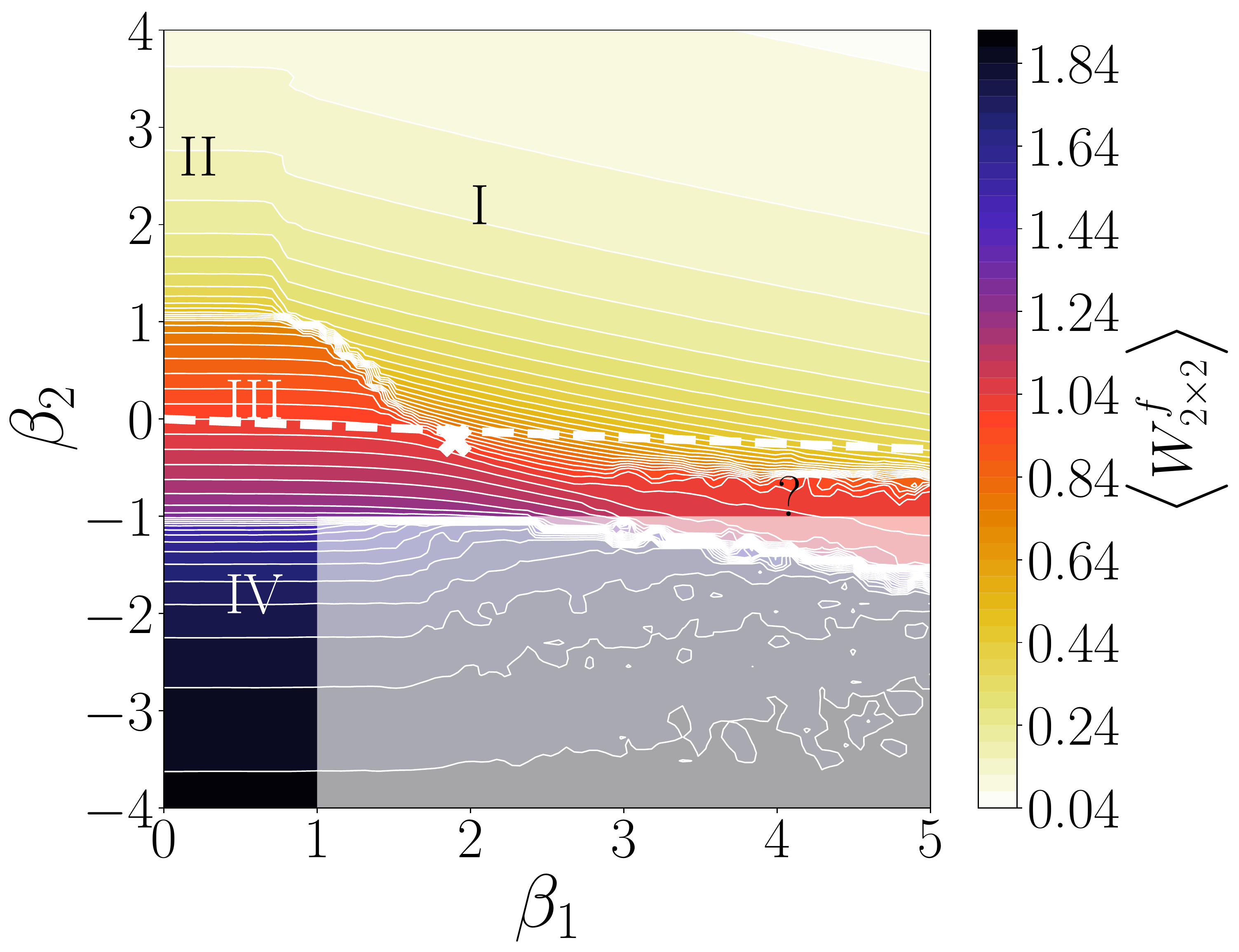}
\par\end{centering}
\caption{Survey of the parameter space, for $L=4$. The image on the \textbf{left}
hand-side shows $\left\langle W_{1\times1}^{f}\right\rangle _{\beta_{1},\beta_{2}}$
and the one on the \textbf{right} hand-side $\left\langle W_{2\times2}^{f}\right\rangle _{\beta_{1},\beta_{2}}$.
The white cross marks the point we selected for subsequent analysis.
We clearly identified four different phases, labeled by roman-numerals.
The thin white lines are level curves. Discontinuities are identified
when several white lines come together. The dashed white line distinguishes
regions with different signs of the coupling, for the naive continuum
limit ($\nicefrac{1}{g^{2}}<0$ below the line, $\nicefrac{1}{g^{2}}>0$
above the line). There might be a new phase in the region around the
question mark (?).\label{fig:PhaseSpaceScan2Loops}}
\end{figure}

\subsection{Variable Size Wilson Loops }

The extension with square Wilson loops of size 1 and 2, given by Eq.\ref{eq:2 loops action},
was selected with the goal of surveying a different region of the
coupling space. We performed a similar analysis as previously, in
which we start by surveying the parameter space to identify the relevant
regions and then we measure the first and second derivative of the
entropy over the regions of interest. The results are shown in Fig.\ref{fig:PhaseSpaceScan2Loops}.
Similarly to the previous extension we can identify several phases,
and there is a single boundary over which we may find a second order phase transition. It is important to point out that
the bottom left corner is washed out since we observed frustration
and were not able to properly perform any measurements.

\begin{figure}[t]
\begin{centering}
\includegraphics[width=0.9\columnwidth]{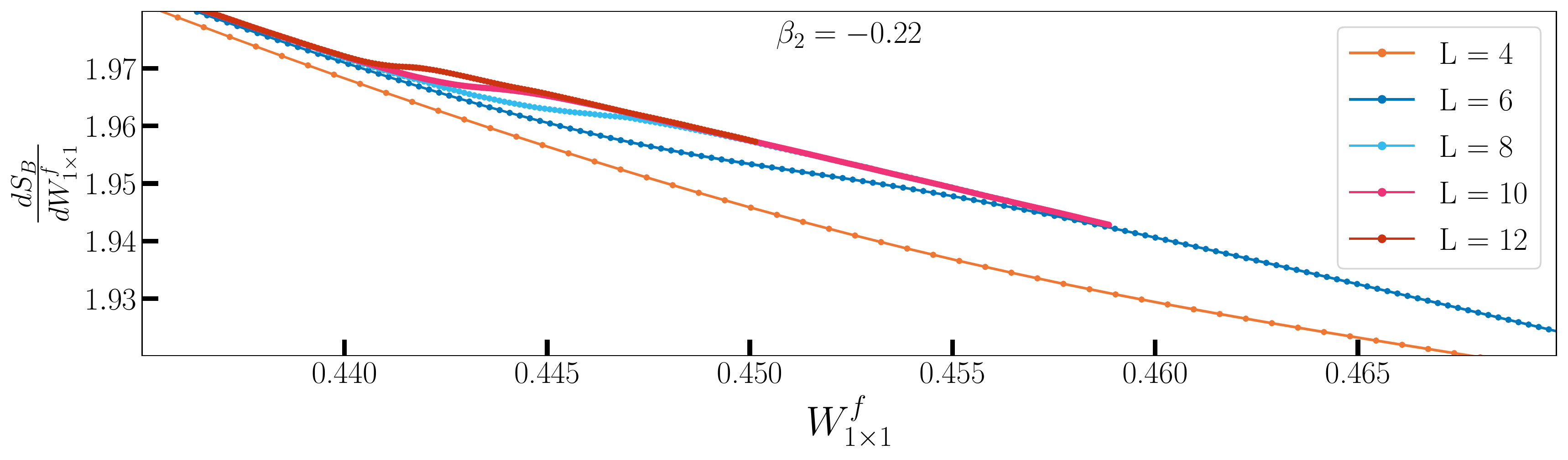}
\par\end{centering}
\begin{centering}
\includegraphics[width=0.9\columnwidth]{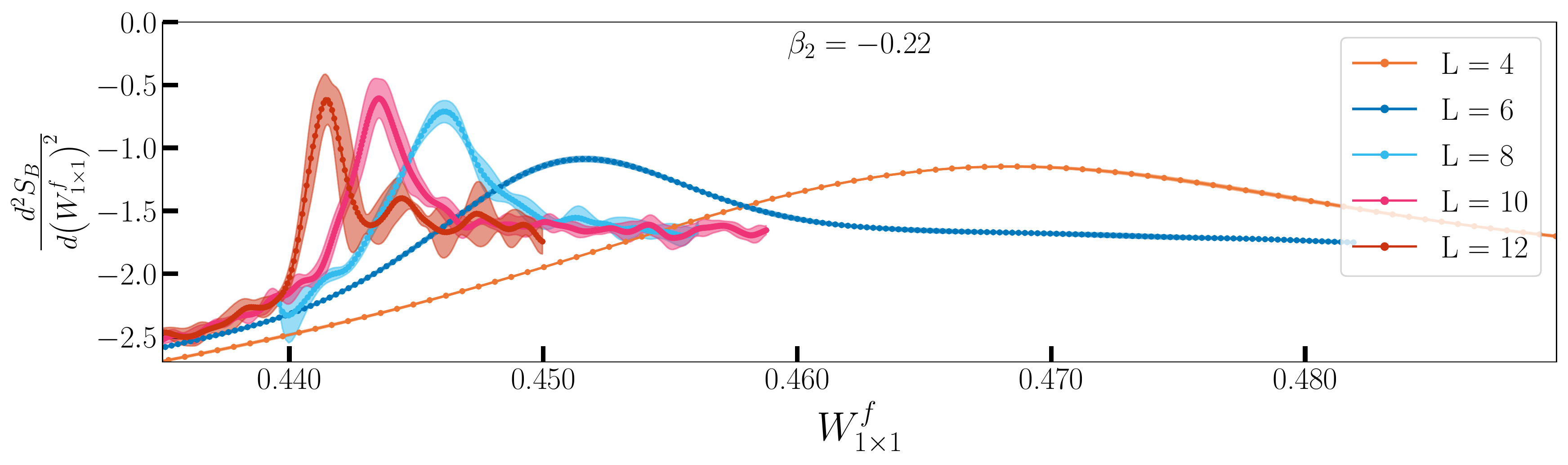}
\par\end{centering}
\centering{}\caption{First (\textbf{top}) and second derivative (\textbf{bottom}) of the
entropy as a function of $W_{1\times1}^{f}$, for different lattice
sizes and $\beta_{2}=-0.22$. The colored region around the markers
is an estimate of the error based on independent measurements. \label{fig:First and Second derivative 2Loops}}
\end{figure}

We performed measurements at several points in the boundary between
region I and III, but we will focus on the results for $\beta_{2}=-0.22$
(signaled by the white cross in Fig.\ref{fig:PhaseSpaceScan2Loops}),
shown in Fig.\ref{fig:First and Second derivative 2Loops}.
Although the observed behavior is similar to the fundamental + adjoint,
there is a significant difference. The first order phase transition
actually disappears over this line, and for $\beta_{2}=-0.22$ we no
longer observe first order criticality. In the bottom plot of Fig.\ref{fig:First and Second derivative 2Loops}
we can see that the maximum of the second derivative is always smaller
than zero, excluding first and second order criticality. 

\begin{figure}[t]
\begin{centering}
\includegraphics[width=0.9\columnwidth]{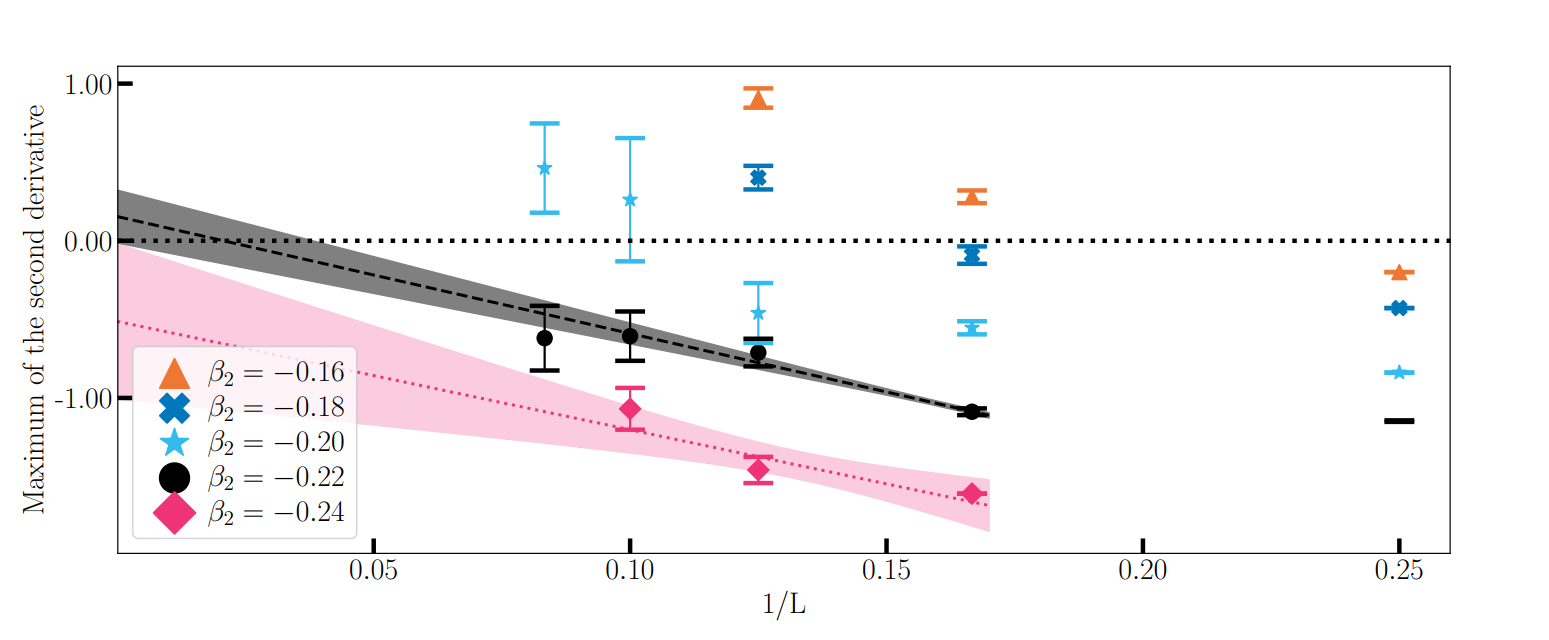}
\par\end{centering}
\caption{Dependence of the maximum of the second derivative of the entropy
as a function of $\nicefrac{1}{L}$, for different $\beta_{2}$ (colors).
The error bars are estimate from independent simulations. \label{fig:Latent Heat}}

\end{figure}

The next step is to try to understand the behavior of this system
when we increase the lattice size. Although the maximum of the second
derivative seems to increase with the lattice size, it is not clear
if it will reach zero, as required for a second order phase transition.
In order the clarify this point we plotted the maximum of the second
derivative of the entropy as a function of the inverse lattice size
for different values of $\beta_{2}$ around $\beta_{2}=-0.22$. The
results, presented in Fig. \ref{fig:Latent Heat}, show that increasing
the lattice size and decreasing $\beta_{2}$ have opposite effects
in the maximum of the second derivative. This allow us to speculate
that there might be a special value
of $\beta_{2}$ such that, when we increase the lattice size, the
maximum of the second derivative goes to zero. To emphasize this point,
we performed a linear fit (dashed lines) for $\beta_{2}=-0.22$ and
$\beta_{2}=-0.24$ and estimated its uncertainty (shaded region).
This fit was made to guide the eye and not as a quantitative extrapolation
of the infinite lattice size limit, as we cannot quantitatively study this limit. 

We conclude that our data is compatible with the existence of second
order criticality for some value of $\beta_{2}$ over the boundary
between region I and III identified in Fig.\ref{fig:PhaseSpaceScan2Loops}. However, as we cannot perform a quantitative study of the infinite lattice size limit, we cannot verify the existence of this fixed point. 

\section{Conclusion}

In this worked we looked for the putative non-trivial CFT in 5-dimensional
pure SU(2) Yang-Mills. This search, motivated by the $\varepsilon$-expansion
\cite{Peskin:1980ay}, was performed using an extension of the lattice
action proposed by \cite{Wilson:1973jj}, where we added to the action terms either
in the adjoint representation (fundamental + adjoint action) or for
square Wilson loops of side 2. We used Monte Carlo methods in which we replaced the
usual canonical ensemble by a generalized ensemble, allowing us to distinguish very weak first order phase transitions from second order phase transitions.

We examined the suggestion made in \cite{Kawai:1992um}, that for
negative enough values of $\beta_{a}$ we might find a second order
phase transition. While this holds true for small lattices ($L=4$),
first order criticality is always recovered for larger lattice sizes
($L\ge6)$, up to at least $\beta_a = -10$. Our analysis shows that, although the critical region
becomes narrower along the boundary between phases I and III, it is
unlikely that first order criticality disappears. 

In the second extension, we observed a behavior compatible with the
existence of a second order phase transition. However, as we are very
limited in the lattice sizes, we could not perform
a quantitative extrapolation of the large lattice size limit, which
would be required to prove the existence of the putative fixed point.
Larger simulations, possible with more sophisticated algorithms, will
be required to definitely answer this question. 

\bibliographystyle{JHEP}
\bibliography{ref,5d_Yang-Mills}

\providecommand{\href}[2]{#2}\begingroup\raggedright\begin{thebibliography}{10}

\bibitem{Florio:2021uoz}
A.~Florio, J.a.M.V.P.~Lopes, J.~Matos and J.a.~Penedones, \emph{{Searching for
  continuous phase transitions in 5D SU(2) lattice gauge theory}},
  \href{https://arxiv.org/abs/2103.15242}{{\ttfamily 2103.15242}}.

\bibitem{Wilson:1973jj}
K.G.~Wilson and J.B.~Kogut, \emph{{The Renormalization group and the epsilon
  expansion}}, \href{https://doi.org/10.1016/0370-1573(74)90023-4}{\emph{Phys.
  Rept.} {\bfseries 12} (1974) 75}.

\bibitem{Peskin:1980ay}
M.E.~Peskin, \emph{{Critical point behavior of the Wilson loop}},
  \href{https://doi.org/10.1016/0370-2693(80)90848-5}{\emph{Phys. Lett. B}
  {\bfseries 94} (1980) 161}.

\bibitem{Kawai:1992um}
H.~Kawai, M.~Nio and Y.~Okamoto, \emph{{On existence of nonrenormalizable field
  theory: Pure SU(2) lattice gauge theory in five-dimensions}},
  \href{https://doi.org/10.1143/PTP.88.341}{\emph{Prog. Theor. Phys.}
  {\bfseries 88} (1992) 341}.

\bibitem{Polyakov:1975rr}
A.M.~Polyakov, \emph{{Interaction of Goldstone Particles in Two-Dimensions.
  Applications to Ferromagnets and Massive Yang-Mills Fields}},
  \href{https://doi.org/10.1016/0370-2693(75)90161-6}{\emph{Phys. Lett. B}
  {\bfseries 59} (1975) 79}.

\bibitem{Bardeen:1976zh}
W.A.~Bardeen, B.W.~Lee and R.E.~Shrock, \emph{{Phase Transition in the
  Nonlinear $\sigma$ Model in 2 + $\epsilon$ Dimensional Continuum}},
  \href{https://doi.org/10.1103/PhysRevD.14.985}{\emph{Phys. Rev. D} {\bfseries
  14} (1976) 985}.

\bibitem{Arefeva:1978fj}
I.Y.~Arefeva, E.R.~Nissimov and S.J.~Pacheva, \emph{{Bphzl Renormalization of
  1/$N$ Expansion and Critical Behavior of the Three-dimensional Chiral
  Field}}, \href{https://doi.org/10.1007/BF01197293}{\emph{Commun. Math. Phys.}
  {\bfseries 71} (1980) 213}.

\bibitem{Eichhorn:2018yfc}
A.~Eichhorn, \emph{{An asymptotically safe guide to quantum gravity and
  matter}}, \href{https://doi.org/10.3389/fspas.2018.00047}{\emph{Front.
  Astron. Space Sci.} {\bfseries 5} (2019) 47}
  [\href{https://arxiv.org/abs/1810.07615}{{\ttfamily 1810.07615}}].

\bibitem{Creutz:1979dw}
M.~Creutz, \emph{{Confinement and the Critical Dimensionality of Space-Time}},
  \href{https://doi.org/10.1103/PhysRevLett.43.553}{\emph{Phys. Rev. Lett.}
  {\bfseries 43} (1979) 553}.

\bibitem{PhysRevE.74.046702}
J.~Viana~Lopes, M.D.~Costa, J.M.B.~Lopes~dos Santos and R.~Toral,
  \emph{Optimized multicanonical simulations: A proposal based on classical
  fluctuation theory},
  \href{https://doi.org/10.1103/PhysRevE.74.046702}{\emph{Phys. Rev. E}
  {\bfseries 74} (2006) 046702}.

\bibitem{Berg:1992qua}
B.A.~Berg and T.~Neuhaus, \emph{{Multicanonical ensemble: A New approach to
  simulate first order phase transitions}},
  \href{https://doi.org/10.1103/PhysRevLett.68.9}{\emph{Phys. Rev. Lett.}
  {\bfseries 68} (1992) 9}
  [\href{https://arxiv.org/abs/hep-lat/9202004}{{\ttfamily hep-lat/9202004}}].

\bibitem{PhysRevLett.71.211}
J.~Lee, \emph{New monte carlo algorithm: Entropic sampling},
  \href{https://doi.org/10.1103/PhysRevLett.71.211}{\emph{Phys. Rev. Lett.}
  {\bfseries 71} (1993) 211}.

\bibitem{PhysRevLett.63.1195}
A.M.~Ferrenberg and R.H.~Swendsen, \emph{Optimized monte carlo data analysis},
  \href{https://doi.org/10.1103/PhysRevLett.63.1195}{\emph{Phys. Rev. Lett.}
  {\bfseries 63} (1989) 1195}.

\bibitem{Creutz:1980wj}
M.~Creutz, \emph{{Asymptotic Freedom Scales}},
  \href{https://doi.org/10.1103/PhysRevLett.45.313}{\emph{Phys. Rev. Lett.}
  {\bfseries 45} (1980) 313}.

\end{thebibliography}\endgroup

\end{document}